\documentclass[conference]{IEEEtran}
\IEEEoverridecommandlockouts

\usepackage{cite}
\usepackage{booktabs}
\usepackage{amsmath,amssymb,amsfonts}
\usepackage{algorithmic}
\usepackage{graphicx}
\usepackage{textcomp}
\usepackage{float}
\usepackage{placeins}

\usepackage{xcolor}
\def\BibTeX{{\rm B\kern-.05em{\sc i\kern-.025em b}\kern-.08em
    T\kern-.1667em\lower.7ex\hbox{E}\kern-.125emX}}

\begin{document}

\title{Explainable Hybrid Feature Selection for Intrusion Detection in Internet of Medical Things Environments}

%
%


\author{
\IEEEauthorblockN{Amira Berrezzek}
\IEEEauthorblockA{
Computer Science Dept.\\
Badji Mokhtar University \\
Annaba, Algeria\\
amira.berrezzek@univ-annaba.dz
}
\vspace{-21pt}
\and
\IEEEauthorblockN{Hayet Djellali}
\IEEEauthorblockA{
Computer Science Dept.\\
Badji Mokhtar University \\
Annaba, Algeria\\
hayet.djellali@univ-annaba.dz
}
\vspace{-21pt}
\and
\IEEEauthorblockN{Giulio Mallardi}
\IEEEauthorblockA{
Computer Science Dept.\\
University of Bari\\
Bari, Italy\\
giulio.mallardi@uniba.it
}
\vspace{-21pt}
\and
\IEEEauthorblockN{Lamia Mahnane}
\IEEEauthorblockA{
Computer Science Dept.\\
Badji Mokhtar University \\
Annaba, Algeria\\
lamia.mahnane@univ-annaba.dz
}
\vspace{-21pt}

}

\maketitle

\begin{abstract}
Internet of Medical Things (IoMT) networks are hard to protect: devices are heterogeneous, computing resources are scarce, and traffic must be analyzed in real time. We present an intrusion detection system that addresses these constraints through feature selection. A Pearson correlation filter first removes redundant attributes; a hybrid strategy then combines model-based feature importance with SHAP attribution to pick a compact subset, on which we train Random Forest and LightGBM classifiers. SHAP and LIME explain what each retained feature contributes to the decisions. On CIC-IoMT 2024 and CIC-IDS 2017, the method cuts the feature space by up to 88\% -- from 40 to as few as 5 features -- and accuracy and F1-score stay within a few points of models trained on all features. Compact, interpretable detectors of this kind are practical candidates for deployment on resource-limited medical networks.

\end{abstract}


\section{Introduction}
The Internet of Medical Things (IoMT) is transforming healthcare by enabling continuous monitoring, remote diagnostics, and data-driven care. At the same time, device heterogeneity, limited resources, and real-time constraints widen the attack surface, making IoMT networks attractive targets for sophisticated cyberattacks~\cite{yacoubi2026ai}. Protecting sensitive medical data exchanged across such infrastructures, therefore, remains a critical challenge. Intrusion Detection Systems (IDS) are a key defense mechanism for identifying malicious traffic. In recent years, machine learning approaches---including ensemble methods such as Random Forest (RF)~\cite{breiman2001random} and LightGBM~\cite{ke2017lightgbm}---have shown strong performance on high-dimensional network data~\cite{berrezzek2025survey}. However, IoMT traffic typically contains many redundant and irrelevant features, which can increase computational cost, reduce generalization, and hinder interpretability~\cite{neto2024review}. Feature selection is thus essential to retain discriminative information while reducing dimensionality.

Beyond accuracy, IoMT security applications increasingly require transparency. Explainable AI (XAI) techniques such as SHAP~\cite{lundberg2017unified} and LIME~\cite{ribeiro2016should} provide global and local insights into model decisions, supporting trust and adoption in safety-critical settings. Nevertheless, many IoMT IDS solutions still treat feature selection and explainability as separate steps, limiting the interpretability of the selected feature subset and the trustworthiness of the resulting models. To address this gap, we propose a feature selection framework in which explainability is part of the selection itself, not an afterthought. A Pearson correlation filter first discards redundant attributes; the surviving features are then ranked twice, once by model-based importance and once by SHAP attribution, and the two rankings are merged into a single compact subset. We train RF and LightGBM on these subsets and use SHAP and LIME to explain the resulting models at the global and local level. Across CIC-IoMT 2024 and CIC-IDS 2017, the subsets shrink to 5--13 features out of 40--70 with little loss in detection performance.

This work makes three contributions:
\begin{itemize}
\item a feature selection pipeline that merges correlation filtering, model-based importance, and SHAP attribution into compact, explainable subsets;
\item experiments with RF and LightGBM on two benchmarks, under several selection strategies, that quantify how much accuracy costs each feature removed;
\item a SHAP and LIME analysis that ties the selected features to the traffic patterns behind intrusion decisions in IoMT networks.
\end{itemize}

Section~II reviews related work, Section~III describes the architecture, Section~IV presents the results, and Section~V concludes.

\section{Background and Methods}
\subsection{Related Work}
Zukaib et al. in~\cite{zukaib2024meta} highlight that the Internet of Medical Things (IoMT) is vulnerable to cyberattacks, requiring advanced intrusion detection. They proposed a meta-learning based Meta-IDS, combining signature-based and anomaly-based techniques. It was evaluated on the WUSTL-EHMS-2020, IoTID20, and WUSTL-IIOT-2021 datasets, achieving up to 99.99\% accuracy.

Following this approach, Salehpour et al. in~\cite{salehpour2025optimized} proposed a resource-efficient IoMT IDS using two-step feature selection with Random Forest classification. It was evaluated on WUSTL-EHMS-2020, NSL-KDD, and CIC-IoMT2024, achieving high accuracy, especially for DDoS/DoS attacks.
In a different direction, Augusta et al. in~\cite{augusta2026blockchain} introduced a Blockchain-enabled IoMT healthcare system with a hybrid IDS using Enhanced Artificial Bee Colony (E-ABC) and Deep Belief Network (DBN). The approach combines optimization with deep learning to enhance detection performance in IoMT settings.
Similarly focused on feature selection and classification, Geetha et al. in~\cite{geetha2024cvsfln} proposed an IoT-IDS using Chaotic Vortex Search (CVS) for feature selection and Fast-Learning Network (FLN) for classification. Evaluated on CIC IDS-2017 and BoT-IoT datasets, it achieved up to 99.7\% accuracy and a 99.81\% detection rate.

In the same context of ensemble learning, Abdullah et al. in~\cite{abdullah2025ensemble} proposed a web intrusion detection model for IoMT using ensemble learning with XGBoost, KNN, Decision Tree, Random Forest~\cite{breiman2001random}, and AdaBoost. Evaluated on medical system traffic data, XGBoost achieved 99.67\% accuracy with \(P<0.001\), while ROC-AUC scores were very high across classifiers. 
On another note, Kumari et al. in~\cite{kumari2025gsa} proposed an IDS using Gravitational Search Algorithm (GSA) for feature selection, Synthetic Minority Over-sampling Technique Iterative Partitioning Filter (SMOTE-IPF) for data balancing, and ML classifiers. Random Forest achieved 99.6\% and 94.4\% accuracy on the NSL-KDD and UNSW-NB15 datasets, respectively.

Similarly, Zeghida et al. in~\cite{zeghida2023mqtt} introduced an IoT intrusion detection system using ensemble learning (bagging, boosting, and stacking) on a balanced MQTT dataset. Compared to single ML models, the approach improved prediction performance, achieving up to 95\% accuracy and F1-score, with MCC exceeding 90\%.
Moving towards hybrid anomaly detection, Zachos et al. in~\cite{zachos2025anomaly} proposed an anomaly-based IDS (AIDS) for IoMT networks using novelty and outlier detection techniques. In contrast to heavier models, results show very low computational overhead, with CPU usage below 1\%, making it suitable for resource-constrained IoMT environments. 

Building on ensemble learning frameworks, Arreche et al. in~\cite{arreche2024two} proposed a two-level ensemble learning framework for network intrusion detection. Following the trend of integrating explainability, it combines multiple ensemble techniques with XAI-based feature selection~\cite{lipton2016mythos}, and was evaluated on RoEduNet-SIMARGL2021, NSL-KDD, and CICIDS-2017 using standard metrics.
Finally, Arreche et al. in~\cite{arreche2024exai} extended this line of research by proposing an end-to-end framework for evaluating black-box XAI methods~\cite{lipton2016mythos} in IDS. Similar to previous studies emphasizing explainability, SHAP and LIME were assessed globally and locally on three benchmark datasets, highlighting the strengths and limitations of current XAI techniques in IDS applications.

Despite these advances, most existing works treat feature selection and explainability as separate processes. Only limited studies integrate explainability directly into the feature selection stage to obtain compact and interpretable feature subsets. This limitation motivates the proposed hybrid explainable feature selection framework, which jointly optimizes detection performance and interpretability for IoMT intrusion detection.

\subsection{Explainable AI and Feature Selection Background}

Feature selection is a crucial preprocessing step that aims to identify the most discriminative attributes by removing irrelevant and redundant variables, thereby improving model efficiency and preserving the most informative features~\cite{neto2024review}. Explainable Artificial Intelligence (XAI) refers to a set of techniques designed to clarify how machine learning models produce their predictions, enabling human users to interpret and understand the reasoning behind the outputs. This is particularly important in high-stakes domains such as healthcare and cybersecurity, where transparency and trust are essential~\cite{lipton2016mythos}.

SHapley Additive exPlanations (SHAP) is a widely used XAI method based on game theory that assigns each feature a contribution score to the model output. It provides both global and local interpretability, enabling the identification of the most influential features~\cite{lundberg2017unified}. Local Interpretable Model-agnostic Explanations (LIME) is a technique that explains individual predictions by approximating the original model locally with an interpretable surrogate model, allowing analysis of the features driving specific intrusion detection decisions~\cite{ribeiro2016should}.

\section{Proposed Hybrid Feature Selection Architecture}
\subsection{Overview of the Proposed Framework}
The framework is organized as a pipeline. Preprocessing handles missing and infinite values and normalizes feature distributions. A Pearson correlation filter ($|r| > 0.9$) then discards highly correlated attributes, so the ranking steps operate on non-redundant features. The hybrid stage scores each surviving feature by model-based importance and by SHAP attribution, and the two scores together decide which attributes are kept. Finally, Random Forest and LightGBM are trained on the resulting subsets, and SHAP and LIME explain the trained models -- SHAP at the level of overall feature contributions, LIME on individual predictions.

\subsection{Datasets}
The CIC-IoMT-2024 dataset~\cite{dadkhah2024cic}, developed by the Canadian Institute for Cybersecurity, is a benchmark for IoMT security research. It contains realistic network traffic from 40 devices (25 real, 15 simulated). The dataset includes benign traffic and 18 simulated attacks, mainly DoS and DDoS~\cite{dadkhah2024cic}. This study focuses on the Wi-Fi and MQTT data. In this work, the dataset is used for binary classification (benign vs. attack) by grouping the considered attack types into a single malicious class. The corresponding files were merged and shuffled, then split into training (70\%) and testing (30\%) sets. The dataset contains 40 features, and no manual feature removal was performed before the feature selection stage.

CIC-IDS 2017~\cite{CICIDS2017} is an intrusion detection dataset containing benign traffic and various network attacks. It includes flow records labeled with timestamps, IPs, ports, protocols, and attack types. Traffic was generated over five days using realistic user profiles and common protocols. Due to its large size (approximately 2,830,540 samples with 79 features), we created a balanced subset of 100,000 samples, including 50,000 benign and 50,000 attack instances. In addition, redundant attributes were removed, resulting in 70 features used in our experiments. The subset was then split into training (70\%) and testing (30\%) sets.

\subsection{Data Pre-Processing}
To prevent data leakage, each dataset was first split into training and test sets. All preprocessing steps, including imputation, normalization, and hybrid feature selection, were performed using only the training data and then consistently applied to the test set. Class labels were encoded in binary form, where 0 denotes benign traffic and 1 denotes attack traffic.

A rigorous data pre-processing pipeline was applied to the CIC-IoMT2024 dataset to ensure data quality and effective model training. First, infinite values were identified and removed, as they can disrupt machine learning algorithms. Next, missing values were handled using median imputation to preserve the data distribution. Finally, Z-score standardization was applied to all numerical features to normalize their scales and ensure balanced model learning.

A streamlined preprocessing pipeline was applied to the CIC-IDS2017 dataset to ensure its suitability for intrusion detection modeling. Selected traffic files were concatenated into a unified dataset for analysis. Missing and infinite values were handled, and irrelevant features were removed to improve data quality and reduce dimensionality, and duplicate records were eliminated. Finally, the dataset was split into training and testing sets, and Min-Max normalization was applied to scale features for stable model training. Since both classifiers are tree-based and invariant to monotonic feature transformations, the different scalers do not affect detection performance; scaling mainly ensures comparable standardized thresholds in the LIME explanations.

\subsection{The proposed Model}
\begin{figure}[!ht]
\centerline{\includegraphics[width=0.35\textwidth]{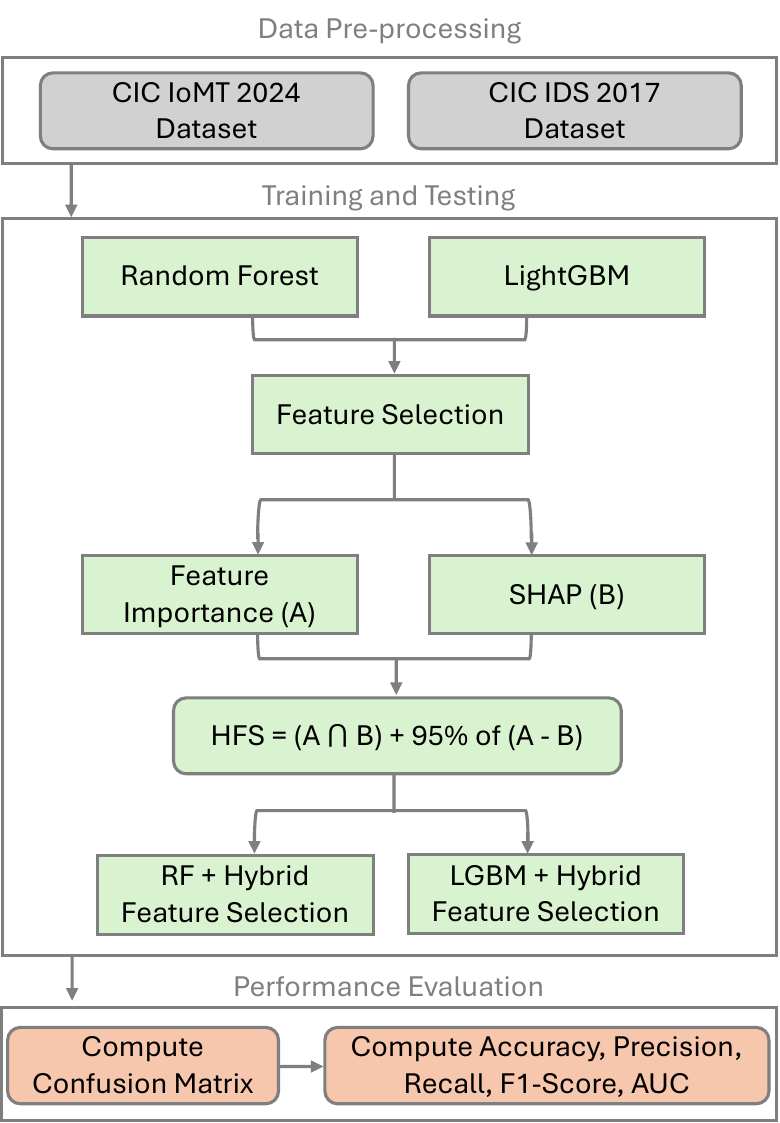}}
\caption{Proposed Architecture}
\label{fig1}
\end{figure}

The core contribution of this work is an explainability-driven feature selection layer that combines model-based feature importance ($A$) with SHAP-based explanations ($B$). The Hybrid Feature Selection (HFS) strategy is defined as:
\begin{equation}
\mathrm{HFS} = (A \cap B) + 95\% \cdot (A - B)
\end{equation}

where $A$ represents the set of features selected using model-based feature importance and $B$ represents the set selected using SHAP importance. The intersection ensures agreement between both methods, while the additional features retained from $A$ preserve predictive information identified by the model. In practice, the $95\%$ threshold corresponds to retaining the top $95\%$ of features according to the Feature Importance ranking used to construct $A$. For example, if Feature Importance initially selects 20 features, the top 19 features are retained. This threshold was empirically determined among candidate values in the 80--100\% range, as it provided the best trade-off between dimensionality reduction and detection performance in preliminary experiments. Random Forest was trained with 100 trees, while LightGBM was trained using its standard configuration. All results were obtained using the same train/test split and preprocessing pipeline described above. The selected features are then used to retrain both models, and performance is evaluated using standard classification metrics and confusion matrices. Fig.~\ref{fig1} illustrates the overall architecture.

\section{Results}
In the following tables, ``+ SHAP'' and ``+ FI'' indicate that the classifier was trained and evaluated using the feature subset selected based on SHAP importance and model-based feature importance, respectively, rather than using SHAP solely for post-hoc explainability.

On the CIC-IoMT 2024 and CIC-IDS 2017 datasets, the results highlight both the effectiveness of Random Forest and LightGBM and the impact of feature selection strategies.

\subsection{CIC-IoMT 2024 Dataset}
Without the Pearson correlation filter, the baseline Random Forest achieves 97.8\% accuracy and F1-score, confirming strong classification performance. Applying SHAP and Feature Importance individually slightly reduces accuracy, while the hybrid approach maintains competitive results (97.2\% accuracy, 99.4\% recall) by combining interpretability and feature relevance into a compact subset. With the Pearson filter, performance further improves to 98.0\% accuracy and F1-score, indicating that removing highly correlated features reduces redundancy and enhances generalization.

\begin{table}[!htbp]

\centering
\caption{Performance Metrics of Random Forest on the CIC IoMT 2024 Dataset, With and Without PC Filtering}
\scriptsize
\begin{tabular}{ c c c c c c }
\hline
 & \multicolumn{5}{c }{Without Pearson Correlation Filter} \\
\hline
 & Accuracy & Precision & Recall & F1-score & AUC \\
\hline
RF & 0.978 & 0.961 & 0.996 & 0.978 & 0.999 \\
\hline
RF + SHAP & 0.975 & 0.960 & 0.991 & 0.975 & 0.999 \\
\hline
RF + FI & 0.971 & 0.950 & 0.994 & 0.971 & 0.999 \\
\hline
Hybrid RF & 0.972 & 0.952 & 0.994 & 0.972 & 0.997 \\
\hline
 & \multicolumn{5}{c }{With Pearson Correlation Filter} \\
\hline
 & Accuracy & Precision & Recall & F1-score & AUC \\
\hline
RF & 0.980 & 0.964 & 0.998 & 0.980 & 0.999 \\
\hline
RF + SHAP & 0.976 & 0.957 & 0.998 & 0.977 & 0.999 \\
\hline
RF + FI & 0.970 & 0.948 & 0.996 & 0.971 & 0.997 \\
\hline
Hybrid RF & 0.978 & 0.985 & 0.971 & 0.978 & 0.996 \\
\hline
\end{tabular}
\label{tab:rf_iomt}
\end{table}
\FloatBarrier
 
\begin{figure}[!ht]
\centering

\includegraphics[width=0.4\textwidth]{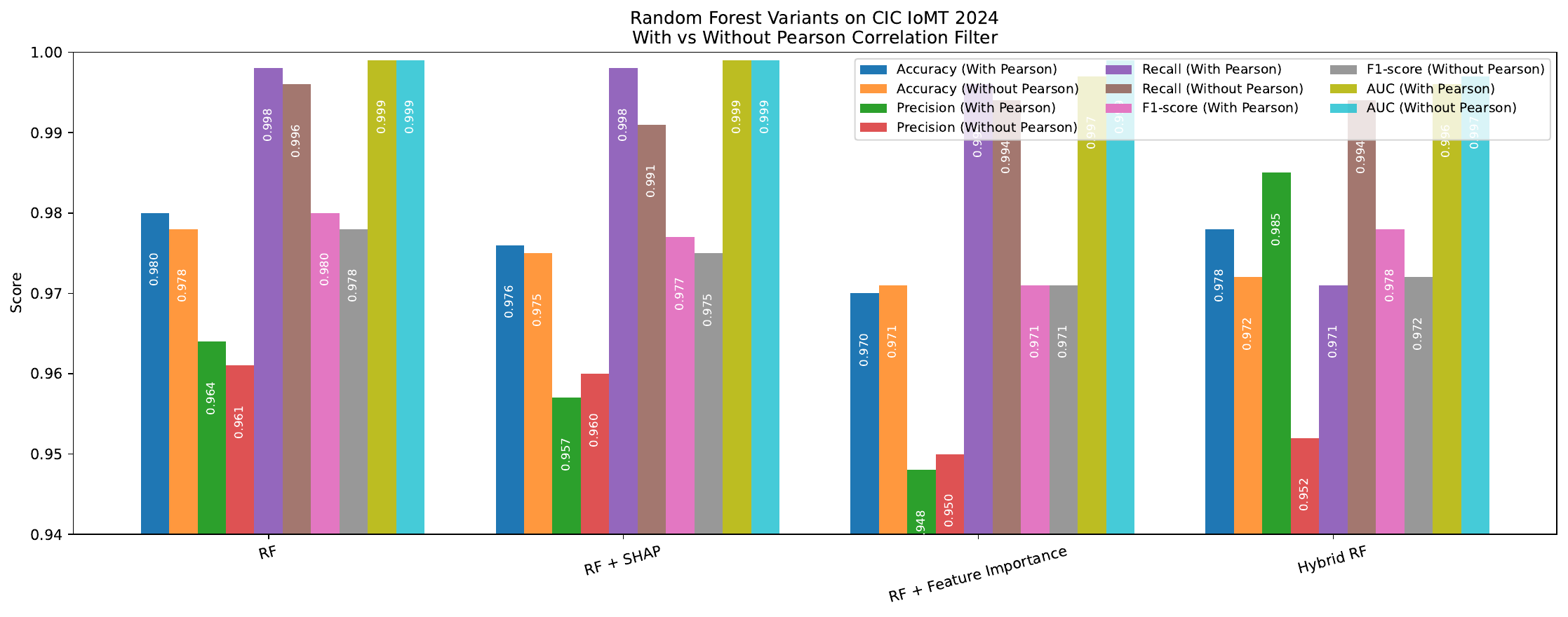}
\caption{Random forest On CIC-IoMT 2024}
\label{fig2}
\end{figure}

\begin{table}[!htbp]
\centering
\caption{Performance Metrics of LightGBM on the CIC-IoMT 2024 Dataset, With and Without PC Filtering}
\scriptsize
\begin{center}
\begin{tabular}{ c c c c c c }
\hline
 & \multicolumn{5}{c }{Without Pearson Correlation Filter} \\
\hline
 & Accuracy & Precision & Recall & F1-score & AUC \\
\hline
LGBM & 0.995 & 0.995 & 0.995 & 0.995 & 0.999 \\
\hline
LGBM + SHAP & 0.960 & 0.929 & 0.995 & 0.961 & 0.997 \\
\hline
LGBM + FI & 0.994 & 0.999 & 0.990 & 0.994 & 0.999 \\
\hline
Hybrid LGBM & 0.960 & 0.928 & 0.998 & 0.961 & 0.998 \\
\hline
 & \multicolumn{5}{c }{With Pearson Correlation Filter} \\
\hline
 & Accuracy & Precision & Recall & F1-score & AUC \\
\hline
LGBM & 0.996 & 0.996 & 0.996 & 0.996 & 0.999 \\
\hline
LGBM + SHAP & 0.962 & 0.931 & 0.997 & 0.963 & 0.999 \\
\hline
LGBM + FI & 0.995 & 0.999 & 0.990 & 0.995 & 0.999 \\
\hline
Hybrid LGBM & 0.985 & 0.976 & 0.995 & 0.986 & 0.999 \\
\hline
\end{tabular}
\label{tab:lgbm_iomt}
\end{center}
\end{table}
 
On the CIC-IoMT 2024 dataset, LightGBM shows excellent performance across all settings. Without Pearson correlation filtering, the baseline LGBM achieves 99.5\% accuracy and F1-score (AUC 99.9\%), while the hybrid method maintains high recall (99.5\%) but lower overall accuracy (96.0\%). 
With correlation filtering, results become more balanced: the baseline reaches 99.6\% accuracy and F1-score, and the hybrid model attains 98.5\% accuracy and 98.6\% F1-score using only 13 of the 40 original features, trading roughly one accuracy point for a 67\% reduction of the feature space.

\begin{figure}[!ht]
\centering
\includegraphics[width=0.4\textwidth]{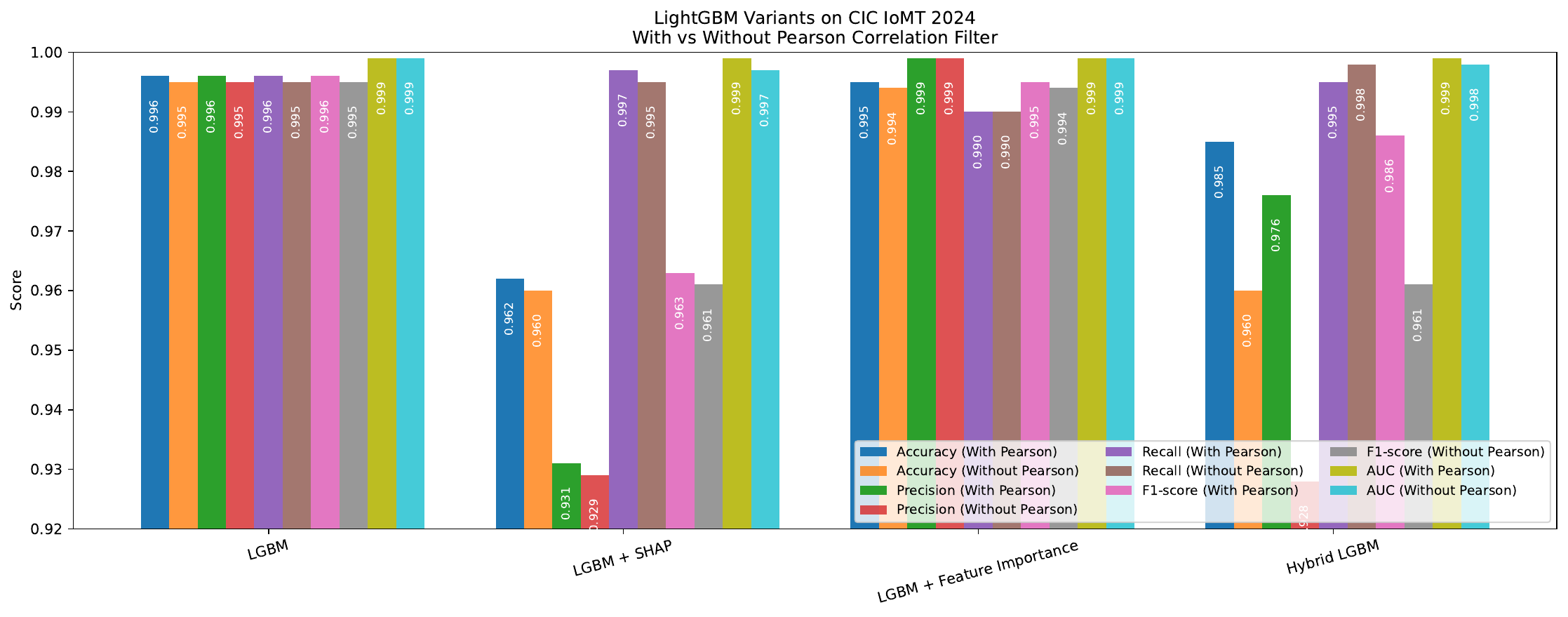}
\caption{LightGBM On CIC-IoMT 2024}
\label{fig3}
\end{figure}
\FloatBarrier
 
\subsection{CIC-IDS 2017 Dataset}
\begin{table}[!htbp]
\centering
\caption{Performance Metrics of Random Forest on the CIC-IDS 2017 Dataset, With and Without PC Filtering}
\scriptsize
\begin{center}
\begin{tabular}{ c c c c c c }
\hline
 & \multicolumn{5}{c }{Without Pearson Correlation Filter} \\
\hline
 & Accuracy & Precision & Recall & F1-score & AUC \\
\hline
RF & 0.998 & 0.999 & 0.998 & 0.998 & 0.999 \\
\hline
RF + SHAP & 0.988 & 0.980 & 0.997 & 0.989 & 0.999 \\
\hline
RF + FI & 0.984 & 0.974 & 0.996 & 0.985 & 0.999 \\
\hline
Hybrid RF & 0.998 & 0.999 & 0.997 & 0.998 & 0.999 \\
\hline
 & \multicolumn{5}{c }{With Pearson Correlation Filter} \\
\hline
 & Accuracy & Precision & Recall & F1-score & AUC \\
\hline
RF & 0.997 & 0.997 & 0.998 & 0.997 & 0.999 \\
\hline
RF + SHAP & 0.985 & 0.975 & 0.994 & 0.985 & 0.994 \\
\hline
RF + FI & 0.999 & 0.999 & 0.999 & 0.999 & 0.999 \\
\hline
Hybrid RF & 0.999 & 0.999 & 0.999 & 0.999 & 0.999 \\
\hline
\end{tabular}
\label{tab:rf_ids2017}
\end{center}
\end{table}
\FloatBarrier
 
On the CIC-IDS 2017 dataset, Random Forest shows excellent performance across all feature selection strategies. The baseline RF achieves 99.8\% accuracy and F1-score, and 99.9\% AUC, while the hybrid approach maintains similar accuracy and F1-score with a high recall of 99.7\%, preserving key features from SHAP and Feature Importance. Applying the Pearson correlation filter further improves consistency, with RF + FI and Hybrid RF reaching up to 99.9\% in all metrics, demonstrating the value of removing redundant features.

\begin{table}[!htbp]
\centering
\caption{Performance Metrics of LightGBM on the CIC-IDS 2017 Dataset, With and Without PC Filtering}
\scriptsize
\begin{center}
\begin{tabular}{ c c c c c c }
\hline
 & \multicolumn{5}{c }{Without Pearson Correlation Filter} \\
\hline
 & Accuracy & Precision & Recall & F1-score & AUC \\
\hline
LGBM & 0.998 & 0.998 & 0.998 & 0.998 & 0.999 \\
\hline
LGBM + SHAP & 0.994 & 0.992 & 0.996 & 0.994 & 0.999 \\
\hline
LGBM + FI & 0.999 & 0.999 & 0.999 & 0.999 & 0.999 \\
\hline
Hybrid LGBM & 0.999 & 0.999 & 0.999 & 0.999 & 0.999 \\
\hline
 & \multicolumn{5}{c }{With Pearson Correlation Filter} \\
\hline
 & Accuracy & Precision & Recall & F1-score & AUC \\
\hline
LGBM & 0.997 & 0.997 & 0.998 & 0.997 & 0.999 \\
\hline
LGBM + SHAP & 0.988 & 0.980 & 0.997 & 0.988 & 0.998 \\
\hline
LGBM + FI & 0.999 & 0.999 & 0.999 & 0.999 & 0.999 \\
\hline
Hybrid LGBM & 0.997 & 0.996 & 0.997 & 0.997 & 0.999 \\
\hline
\end{tabular}
\label{tab:lgbm_ids2017_metrics}
\end{center}
\end{table}
\FloatBarrier

On the CIC-IDS 2017 dataset, LGBM shows excellent intrusion detection performance. Baseline LGBM reaches 99.8\% accuracy, precision, recall, and F1-score. SHAP slightly lowers performance, while Feature Importance and Hybrid LGBM maintain near-optimal results. With the Pearson filter, LGBM + FI stays at 99.9\%, and Hybrid LGBM remains strong (99.7\%), balancing performance, feature reduction, and explainability.

\begin{figure}[!ht]
\centering
\includegraphics[width=0.4\textwidth]{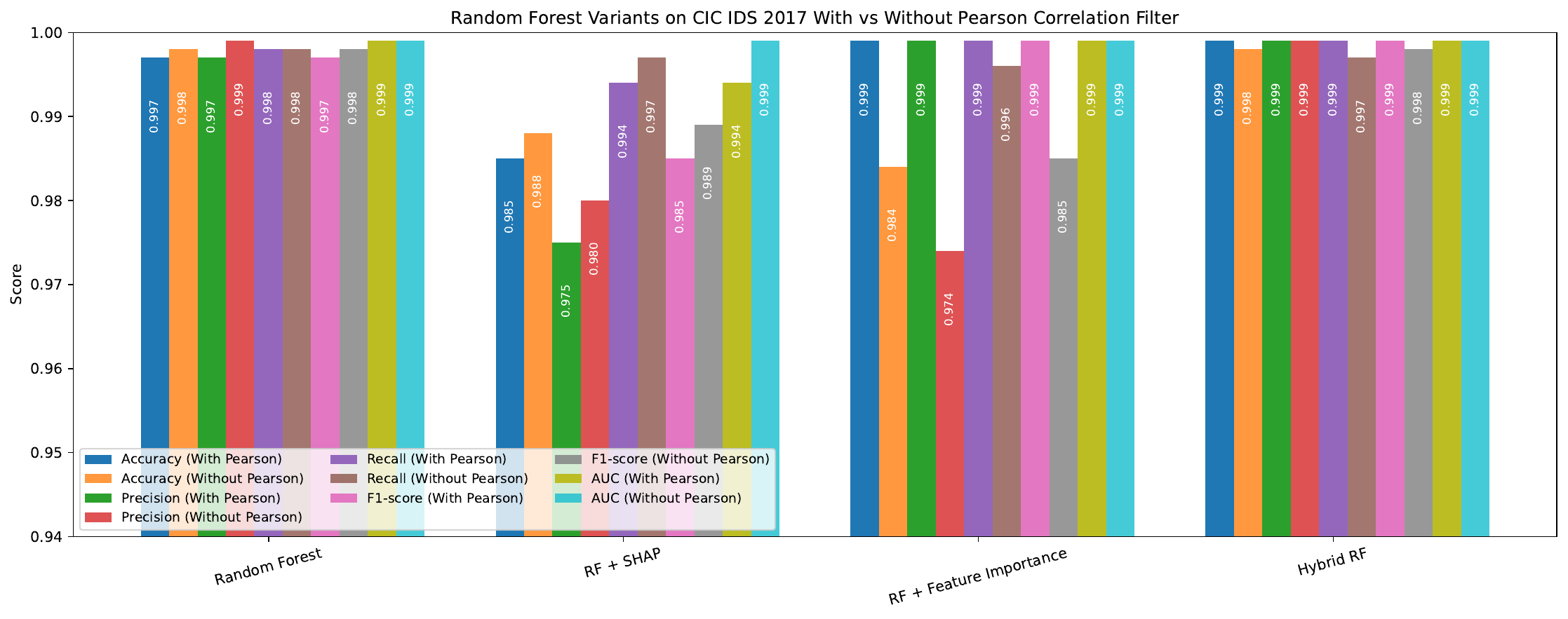}
\caption{Random Forest On CIC-IDS 2017}
\label{fig4}
\end{figure}
\FloatBarrier

\begin{figure}[!ht]
\centering
\includegraphics[width=0.4\textwidth]{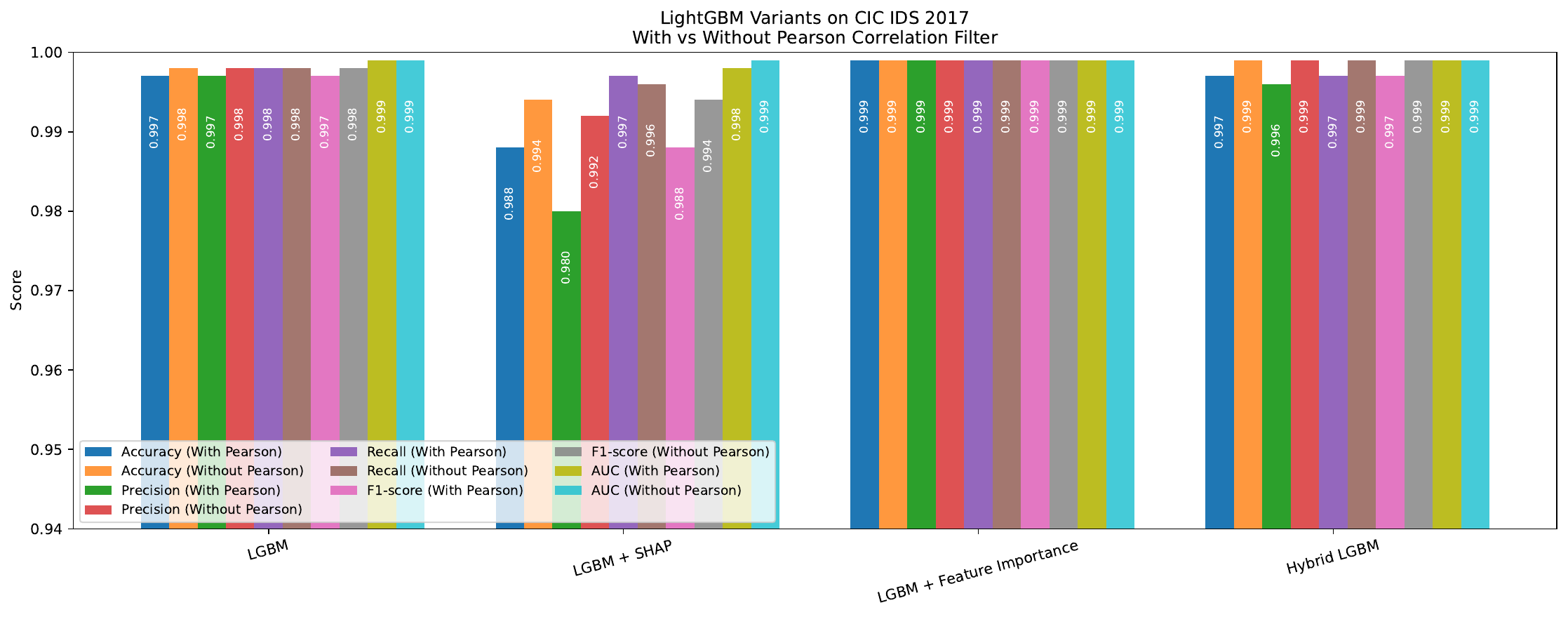}
\caption{LightGBM On CIC-IDS 2017}
\label{fig5}
\end{figure}
\FloatBarrier
 
\subsection{Feature Selection and Explainability Results}
\begin{table}[!htbp]
\centering
\caption{Number of Features Selected by the Proposed Hybrid Feature Selection Method}
\begin{center}
\begin{tabular}{lcc}
\hline
 & CIC-IoMT 2024 & CIC-IDS 2017 \\
\hline
Hybrid RF   & 5  & 12 \\
Hybrid LGBM & 13 & 8  \\
\hline
\end{tabular}
\label{tab:hfs_feature_counts}
\end{center}
\end{table}
\FloatBarrier

These results correspond to a substantial reduction of the original feature space while maintaining high detection performance.
 
\begin{figure}[!ht]
\centering
\includegraphics[width=0.4\textwidth]{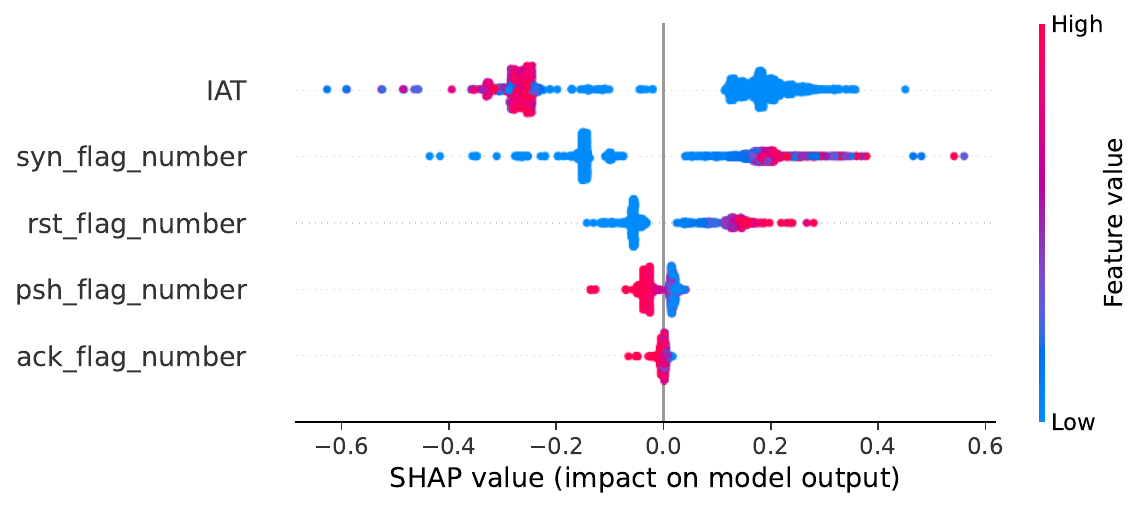}
\caption{Feature Impact Analysis of RF on the CIC IoMT 2024 Dataset}
\label{fig6}
\end{figure}
\FloatBarrier

\begin{table}[!htbp]
\centering
\caption{LIME Explanation for Random Forest on the CIC IoMT 2024 Dataset}
\begin{center}
\begin{tabular}{l l r}
\hline
\FloatBarrier
Feature & Condition & Value \\
\hline
IAT & $\leq -0.83$ & 0.562935 \\
syn flag number & $\leq -0.73$ & -0.015965 \\
psh flag number & $\leq -1.29$ & 0.006832 \\
rst flag number & $\leq -0.68$ & -0.006471 \\
ack flag number & $> 0.80$ & 0.001241 \\
\hline
\end{tabular}
\label{tab:lime_rf_iomt2024}
\end{center}
\end{table}
\FloatBarrier
 
These LIME results indicate that IAT (Inter-Arrival Time) is the most critical predictor, where low values ($IAT \leq -0.83$) strongly push the Random Forest toward a specific classification (likely Malicious) with a high local weight of 0.563. Other features like syn and rst flag counts show negative contributions, while ack flags over a threshold provide a marginal positive influence, collectively defining the model's decision boundary for IoMT traffic.

\begin{figure}[!ht]
\centering
\includegraphics[width=0.4\textwidth]{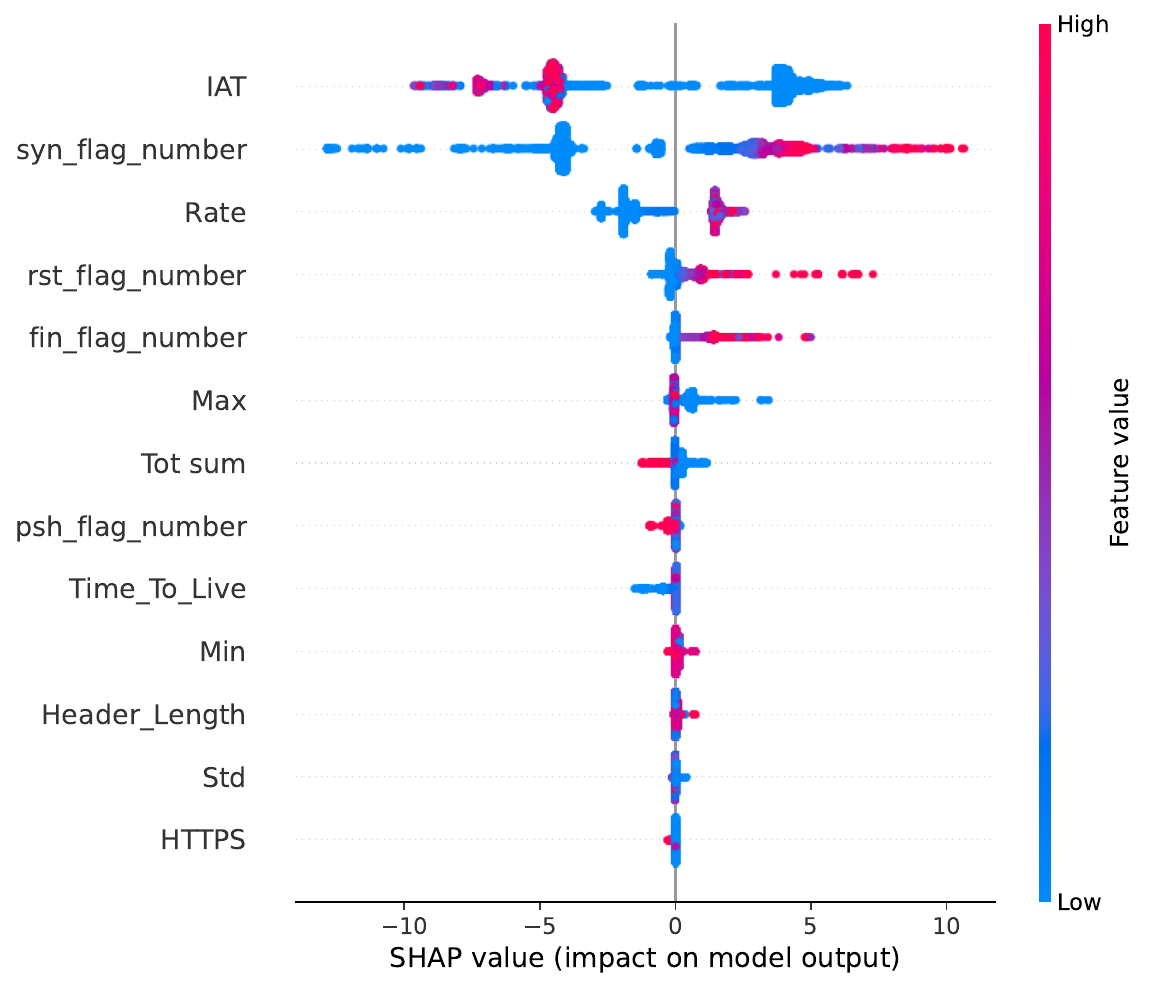}
\caption{Feature Impact Analysis of LGBM on the CIC IoMT 2024 Dataset}
\label{fig7}
\end{figure}
\FloatBarrier
 
\begin{table}[!htbp]
\centering
\caption{LIME Explanation for LGBM on the CIC IoMT 2024 Dataset}
\begin{center}
\begin{tabular}{l l r}
\hline
Feature & Condition & Value \\
\hline
IAT & $\leq -0.83$ & 0.474506 \\
Rate & $> 0.36$ & 0.169202 \\
fin flag number & $\leq -0.52$ & -0.053231 \\
Std & $\leq -0.59$ & 0.030717 \\
Max & $\leq -0.88$ & 0.029507 \\
\hline
\end{tabular}
\label{tab:lime_lgbm_iomt2024}
\end{center}
\end{table}
\FloatBarrier
 
The LIME local explanation for the LightGBM model on the CIC IoMT 2024 dataset reveals that the Inter-Arrival Time (IAT) and traffic Rate are the primary drivers for this specific classification, where an IAT $\leq-0.83$ yields the highest contribution weight ($0.474506$) toward the prediction. This suggests that the model identifies rapid, high-frequency packet patterns characterized by low inter-arrival times and high rates as the critical diagnostic signatures for distinguishing malicious IoMT anomalies from benign medical device telemetry. This signature is consistent with the flooding behavior of the DoS/DDoS attacks that dominate the dataset~\cite{dadkhah2024cic}.

\begin{figure}[!ht]
\centering
\includegraphics[width=0.3\textwidth]{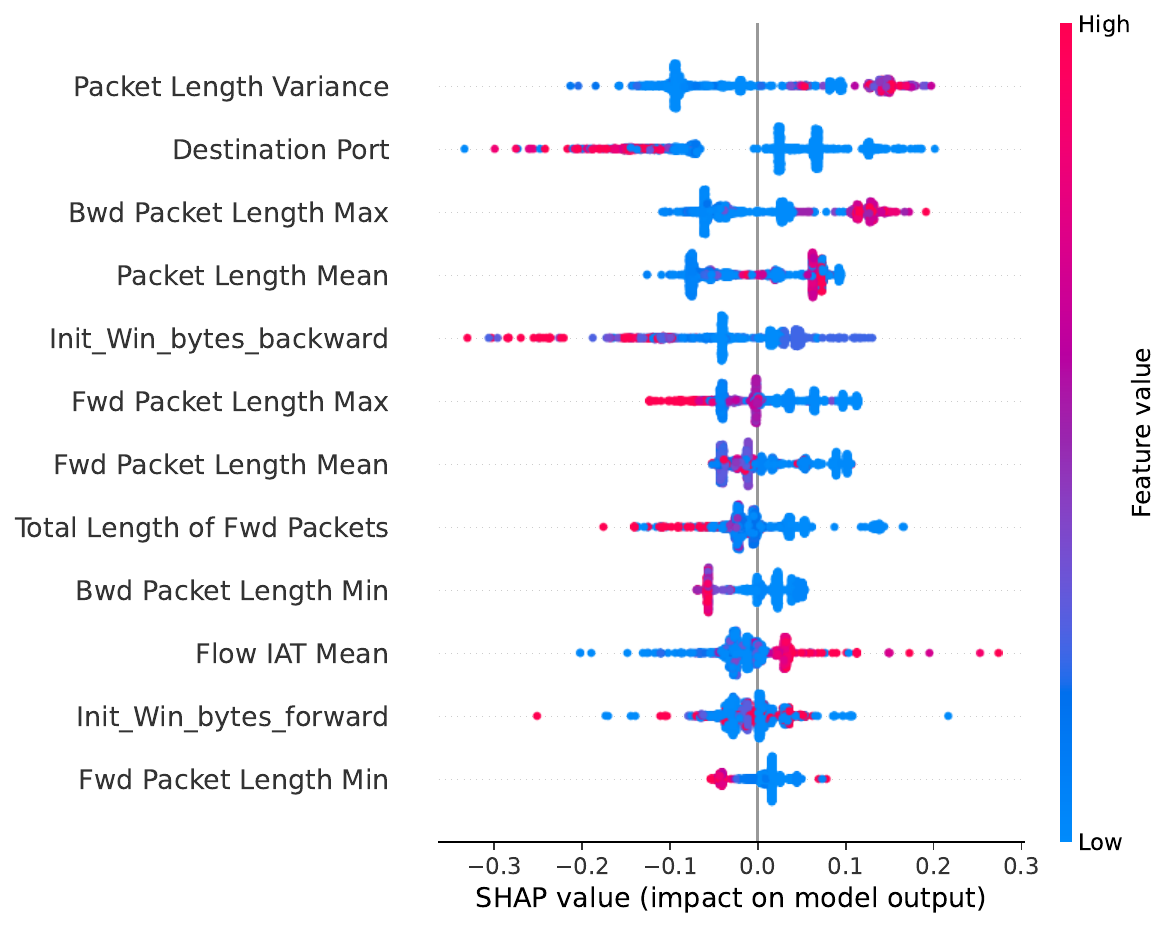}
\caption{Feature Impact Analysis of RF on the CIC IDS 2017 Dataset}
\label{fig8}
\end{figure}
\FloatBarrier

For the CIC IDS 2017 dataset with Random Forest, LIME identifies Destination Port and Packet Length Variance as the dominant localized features, where specific port targeting and high payload irregularity serve as high-weight indicators ($>0.10$) for detecting sophisticated network intrusions.

\begin{table}[!htbp]
\centering
\caption{LIME Explanation for RF on the CIC IDS 2017 Dataset}
\begin{center}
\begin{tabular}{l l r}
\hline
\FloatBarrier
Feature & Condition & Value \\
\hline
Destination Port & $\leq 0.00$ & 0.198658 \\
Packet Length Variance & $> 0.10$ & 0.107942 \\
Init Win bytes backward & $\leq 0.00$ & 0.040776 \\
Bwd Packet Length Max & $> 0.25$ & 0.032947 \\
Bwd Packet Length Min & $\leq 0.00$ & 0.031143 \\
\hline
\end{tabular}
\label{tab:lime_rf_ids2017}
\end{center}
\end{table}
\FloatBarrier

\begin{figure}[!ht]
\centering
\includegraphics[width=0.4\textwidth]{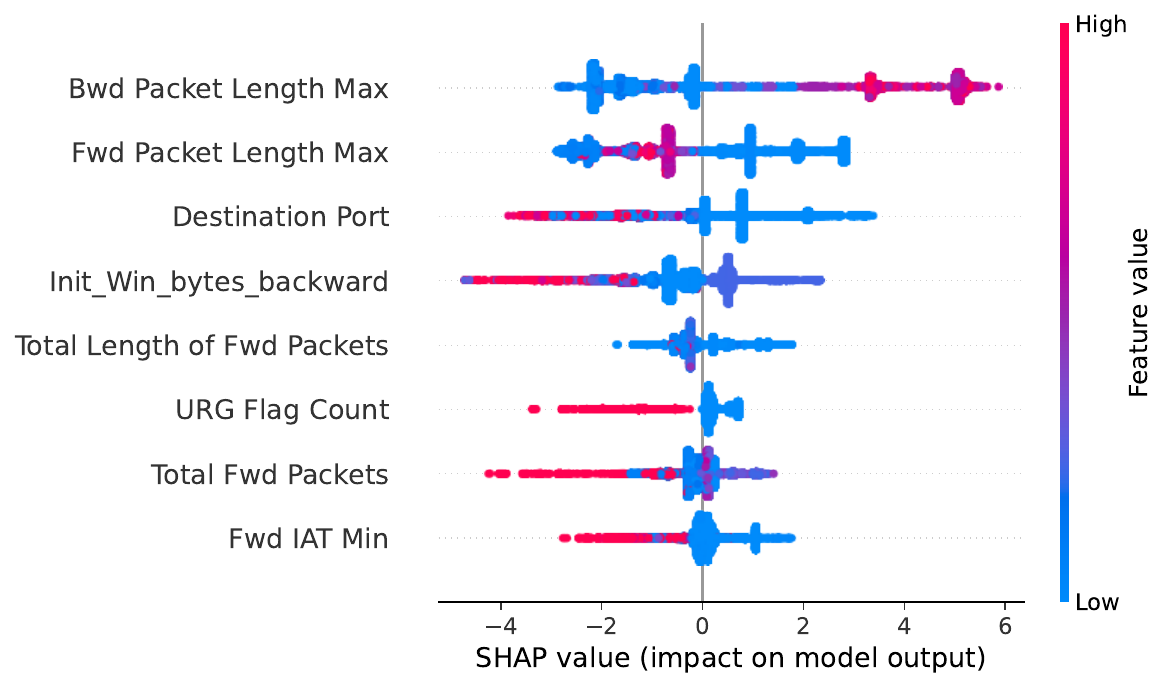}
\caption{Feature Impact Analysis of LGBM on the CIC IDS 2017 Dataset}
\label{fig9}
\end{figure}
\FloatBarrier
 
\begin{table}[!htbp]
\centering
\caption{LIME Explanation for LGBM on the CIC IDS 2017 Dataset}
\begin{center}
\begin{tabular}{l l r}
\hline
\FloatBarrier
Feature & Condition & Value \\
\hline
Destination Port & $\leq 0.00$ & 0.294631 \\
Bwd Packet Length Max & $> 0.25$ & 0.112012 \\
URG Flag Count & $\leq 0.00$ & 0.060921 \\
Fwd Packet Length Max & $\leq 0.00$ & -0.045540 \\
Fwd IAT Min & $\leq 0.00$ & 0.034684 \\
\hline
\end{tabular}
\label{tab:lime_lgbm_ids2017}
\end{center}
\end{table}
\FloatBarrier

LIME confirms that LightGBM prioritizes the Destination Port and Bwd Packet Length Max as the most influential predictors, where specific port targeting and significant backward packet size variability provide a combined contribution weight exceeding $0.40$ toward the final classification. This indicates that the model detects intrusions by identifying specific service layer patterns and traffic volume asymmetries in the backward flow, in line with the port-specific attacks (e.g., FTP/SSH brute force and web attacks) included in CIC-IDS 2017~\cite{CICIDS2017}.
\vspace{-10pt}
\section{Conclusion}
The hybrid feature selection method developed in this work, combined with XAI techniques, reduces the feature space of IoMT intrusion detection models by up to 88\% while keeping detection performance within a few points of full-feature baselines, yielding compact, interpretable models suited to resource-constrained healthcare environments. Beyond efficiency, the integration of SHAP and LIME provides actionable insights into the traffic characteristics driving predictions, supporting transparent and trustworthy defenses for critical healthcare infrastructure. The very high scores observed are consistent with the literature on these benchmarks in binary settings; although leakage was prevented by fitting all preprocessing on training data only, generalization to live, heterogeneous IoMT traffic remains to be assessed. Future work will investigate multiclass intrusion detection scenarios, systematic sensitivity analysis of the selection threshold, and runtime evaluation in real-time IoMT environments.
\bibliographystyle{IEEEtran}
\bibliography{bib}
\end{document}